\documentclass[copyright]{eptcs}

\usepackage{tikz}
\usetikzlibrary{arrows,automata}
\usetikzlibrary{shapes,snakes,backgrounds,petri,calc}
\usetikzlibrary{positioning}
\usepackage[ruled,slide, vlined]{algorithm2e}
\usepackage{multicol}
\usepackage{subfig}
\usepackage{xspace}

\newcommand{\ordi}[3]{\draw[#3] ( #1, #2 ) -- ( #1+1, #2 ) -- ( #1+1, #2+1 )  -- ( #1, #2+1 ) -- ( #1, #2) -- ( #1-.3, #2-.3) -- ( #1+.7, #2-.3 ) -- ( #1+1, #2 ) ;}

\newcommand{\bonhomme}[3] {%
\draw [#3] ( #1, #2 ) circle ( .3 );%
\draw [#3] ( #1, #2-.3 ) -- ( #1, #2-1 );%
\draw [#3] ( #1-.3, #2-.6 ) -- ( #1+.3, #2-.6);%
\draw [#3] ( #1, #2-1 ) -- ( #1-.3, #2-1.6 );%
\draw [#3] ( #1, #2-1 ) -- ( #1+.3, #2-1.6 );%
}

\newcommand{\server}[3]{%
\draw[#3] ( #1, #2) -- ( #1, #2-2 ) -- ( #1+1, #2-2 ) -- ( #1+1, #2 ) -- ( #1, #2 ) -- ( #1+.3, #2+.3 ) -- ( #1+1.3, #2+.3 ) -- ( #1+1.3, #2-1.7) -- ( #1+1, #2-2 );%
\draw[#3] ( #1+1, #2 ) -- ( #1+1.3, #2+.3 ) ;%
}

\newcommand{\ie}{\textit{i.e.}\xspace}

\newcommand{\ignore}[1]{{}}

\title{Parametric, Probabilistic, Timed Resource Discovery System}
\author{Camille Coti
\institute{LIPN, CNRS UMR 7030, Universit\'e Paris 13, Sorbonne Paris Cit\'e}
\email{camille.cot@lipn.univ-paris13.fr}
}
\def\titlerunning{Parametric, Probabilistic, Timed Resource Discovery System}
\def\authorrunning{C. Coti}

\begin{document}
\maketitle

\begin{abstract}
  This paper presents a fully distributed resource discovery and
  reservation system. Verification of such a system is important to
  ensure the execution of distributed applications on a set of
  resources in appropriate conditions. A semi-formal model for his
  system is presented using probabilistic timed automata. This model
  is timed, parametric and probabilistic, making it a challenge to the
  parameter synthesis community.
\end{abstract}

\section{Introduction}

The behavior of distributed systems in general can be challenging to
prove. On the other hand, model checking techniques are particularly
well adapted to verify that they follow a certain specification,
because such techniques explore every possible execution of the
system. Moreover, the execution of a distributed algorithm can depend
on a large number of parameters, because of the complex combination of
elements involved. It is therefore difficult to have a quantitative
evaluation of what happens in the system for each and every value of
each parameter. 

This paper focuses on a resource management system. For some
applications, resources can be shared between clients. These 
clients may need to access the resources in exclusive mode. As a
consequence, a reservation system is necessary to arbitrate between
clients and orchestrate the utilization of the resources. 

For instance, a laboratory can buy a set of computation nodes and put
them together in a cluster. A specific node, called the front-end, is
used to access the computation nodes and a batch scheduler installed
on this front-end node issues reservations on the nodes. The nodes
cannot be accessed by a user that does not have an ongoing reservation
on the said node issued by the batch scheduler. In practice, this
reservation system is implemented using a single job queue that
maintains a list of jobs that need to be executed and a list of
available resources, and schedules the former on the latter
\cite{LSF,OAR,torque,PBS}.

However, in many situations, this architecture cannot be
implemented. For instance, a small lab who bought a handful of GPU
nodes might not want to dedicate hardware and human time to setup and
administrate a front-end node with a batch scheduling system. On some
critical systems, this component can be seen as a single point of
failure and then reduce the reliability of the whole system. However,
in these situations, it is still necessary to make sure that
applications have exclusive access to the machines they are using.

In this paper, we present a completely distributed reservation
system, based on the state maintained by each machine itself, and the
local network protocol Zeroconf \cite{zeroconf}. This system was
modeled using (timed) Petri Nets in \cite{arxivmodel, CEK15}. However,
we have seen that this system is highly parametrizable. the purpose of 
this paper is to present it as a parametric system for verification
and parameter synthesis. The resulting model has a high level of
complexity, and therefore forms a case-study which cannot be
reasonably solved with current methods. With this paper we aim at
presenting it in order to open a discussion for methods that would
tackle these problems, for instance by combining or hybriding
methods. 

The rest of this paper is organized as follow. The global architecture
of the system and the algorithms are described in Section
\ref{sec:presentation}. Section \ref{sec:model} describes how it can
be modeled. Section \ref{sec:params} specifies the expected behavior
of the system and how parameter synthesis is useful to verify this
behavior. Finally, section \ref{sec:conclu} concludes this paper and
opens questions for the community. 

\section{Presentation of the system}
\label{sec:presentation}

This section presents the reservation system itself. The global
architecture is described in section \ref{sec:archi}, the algorithms
are presented in section \ref{sec:reservation} for the reservation
protocol itself and in section \ref{sec:machines} for the machines.  

\subsection{Architecture}
\label{sec:archi}

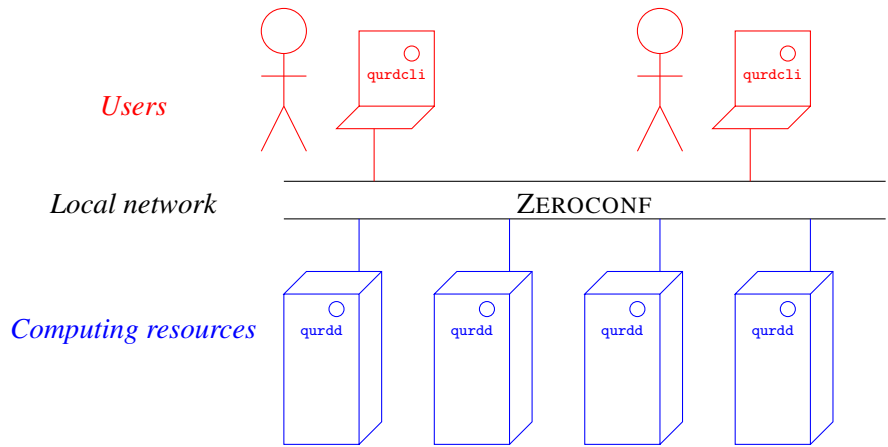
\begin{figure}
  \begin{center}
    \begin{tikzpicture}


\draw ( 0, 0 ) -- ( 8, 0 );
\draw ( 0, -.5 ) -- ( 8, -.5 );
\node at ( 4, -.3) {Z{\sc eroconf}};
\node at ( -2, -.3) {\it Local network};


\ordi{1}{1}{color=red}
\draw [color=red] ( 1.2, .7 ) -- ( 1.2, 0 );
\bonhomme{0}{2}{color=red}
\draw [color=red] ( 1.7, 1.7 ) circle ( .1 );
\node[color=red] at ( 1.5, 1.4) {\tiny \tt qurdcli};

\ordi{6}{1}{color=red}
\draw [color=red] ( 6.2, .7 ) -- ( 6.2, 0 );
\bonhomme{5}{2}{color=red}
\draw [color=red] ( 6.7, 1.7 ) circle ( .1 );
\node[color=red] at ( 6.5, 1.4) {\tiny \tt qurdcli};

\node [color=red] at ( -2, 1) {\it Users};


\server{0}{-1.5}{color=blue}
\draw [color=blue] ( 1, -1.2 ) -- ( 1, -.5 );
\draw [color=blue] ( .7, -1.7 ) circle ( .1 );
\node[color=blue] at ( .5, -2) {\tiny \tt qurdd};

\server{2}{-1.5}{color=blue}
\draw [color=blue] ( 3, -1.2 ) -- ( 3, -.5 );
\draw [color=blue] ( 2.7, -1.7 ) circle ( .1 );
\node[color=blue] at ( 2.5, -2) {\tiny \tt qurdd};

\server{4}{-1.5}{color=blue}
\draw [color=blue] ( 5, -1.2 ) -- ( 5, -.5 );
\draw [color=blue] ( 4.7, -1.7 ) circle ( .1 );
\node[color=blue] at ( 4.5, -2) {\tiny \tt qurdd};

\server{6}{-1.5}{color=blue}
\draw [color=blue] ( 7, -1.2 ) -- ( 7, -.5 );
\draw [color=blue] ( 6.7, -1.7 ) circle ( .1 );
\node[color=blue] at ( 6.5, -2) {\tiny \tt qurdd};

\node [color=blue] at ( -2, -2) {\it Computing resources};

\end{tikzpicture}
    \caption{\label{fig:archi}Architecture of a QURD system.}
  \end{center}
\end{figure}

The global architecture of the system, named QURD, is depicted in
Figure \ref{fig:archi}. Computing resources \emph{declare themselves}
on the Zeroconf bus, and users/clients look on the Zeroconf bus to see
which resources are available. An \emph{application} (or a \emph{job})
is made of several \emph{processes} that are meant to run on a set of
\emph{resources}, also called \emph{machines}. The user submits an
application through a \emph{client}. 

The Zeroconf bus \cite{zeroconf} is a network protocol used for
self-configuration of network services. It was originally designed to
allow automatic configuration of computers without any intervention
from the user nor any centralized server, and later extended to
various services. For instance, it is widely used for services such
as DNS or network printers. In the latter example, printers declare
themselves on the Zeroconf bus, and workstations look on the Zeroconf
bus to find out which printers are available.

Zeroconf uses multicast UDP datagrams. It features three operations,
two of them can be used for automatic service detection:
\emph{discover} and \emph{advertise}. The third operation is called
\emph{resolution}: for instance, it is used by the multicast DNS
protocol (mDNS) works as follows: 1) the client sends a multicast
datagram to ask ``what is the IP address that corresponds to this
symbolic name''; 2) the hosts that has this symbolic name name answers
with a unicast message to the client. A host that provides a service
would typically \emph{advertise} it. For this purpose, it sends a
multicast datagram to tell all the machines of the networks ``I
provide this service and on that port''. The other mode that can be
used for service detection, \emph{discover}, works the other way
around: 1) a client sends a multicast datagram to ask ``who provides
this service?''; 2) servers that host the requested service reply to
the client using a unicast datagram. 

In our system, machines declare themselves on the Zeroconf bus when
they are available and withdraw themselves when they are taken by a
client, \ie when a job is running on them: they use the
\emph{advertise} mode. Clients listen and receive the multicast
declarations that are sent on the network. 

However, this is not sufficient to ensure exclusive access to the
machines, because of the asynchronous nature of the system and the
absence of a consistent view between concurrent clients. Besides, the
Zeroconf protocol does not have real-time accuracy: a machine which is
not available anymore might be still visible on it. For instance, if a
resource is available at a given moment, two clients will see it on
the Zeroconf bus. Both will issue a request, but only one of them should
get it.

\subsection{Reservation algorithm}
\label{sec:reservation}

When a client wants to reserve a set of machines, the submission
protocol works as follows. The client listens to the Zeroconf bus for
available machines. It gets a list of available machines. It contacts
them one by one and waits for a reply. If the machine acknowledges the
reservation, the client receives an ``OK'' message and keeps the
machine. Otherwise, the client receives a ``KO'' message.

Once the client has enough machines, it starts its job on the said
machines. Otherwise, two behaviors are possible. Either the client
waits for other machines to become available, with a limit of time set
by a timeout (called the \emph{wait semantics}), or it cancels its
reservation and frees the machine it has reserved, to try again later
(called the \emph{fail semantics}). The algorithm corresponding to the
latter approach is given by algorithm \ref{algo:reservationFail}.

\begin{algorithm}[ht!b]
  \SetFuncSty{textbf}
  \SetKwFunction{reservenodes}{reserveNodes}
  \reservenodes{ nbNodes } 
  \Begin{
      \KwData{machines = \{\}}
      listenZeroconf()\;
      \ForEach{machine $m$ newly discovered} {
        \If{ card( machines ) $<$ nbNodes } {
          contactMachine( $m$ ) \;
          ack = receiveAck( $m$ )\;
          \If{ ack == OK } {
            machines.append( $m$ ) \;
          }
        }
      }
      \eIf{ card( machines ) == nbNodes } {
        \Return{machines} \;
      } {
        freeMachines( machines ) \;
        \Return{\{\}} \;
      }
      
    }
    \caption{\label{algo:reservationFail}Resource reservation algorithm (fail semantics)}
  \end{algorithm}

One important issue with these two reservation semantics is to avoid
deadlocks between concurrent reservation requests. For instance, if
we consider a system with three machines and two concurrent
reservation requests, one for two machines and one for three
machines. If the first request gets one machine and the second gets
two machines, neither of them has obtained enough machines and there
is no spare machine left. This situation would lead to a deadlock
without the possibility to release and retry (\emph{fail} semantics)
or release everything after a timeout (\emph{wait} semantics). The
\emph{wait} semantics is still useful if there is no available
resource because some of them are used by running applications. In
this case, as soon as an application is done, the request can get its
resources. 
  
\subsection{Exclusive access protocol}
\label{sec:machines}

In order to make sure that jobs have exclusive access to the machines
that have been assigned to them, each machine implements a protocol
based on its state. When a machine is available for jobs, its state
is set to \emph{available}. It can be reserved only when it is in the
stat \emph{available}. Otherwise, in any other state, it answers all
the requests with ``KO''.

\subsection{End of a job}
\label{sec:end}

Once the process running on a machine is done, the machine's state is
set to \emph{finished}. A job ends when all the processes are done:
the client keeps track of which machine is done. Once the job is
finished, the machines switch back into state \emph{available} and
republish themselves on the Zeroconf bus.

\subsection{Resource volatility}
\label{sec:volatility}

Resources can be subject to failures. If they fail while they are
idle, they switch to state \emph{unavailable} or, if they are still
visible on the Zeroconf bus, they never answer the clients'
requests and the clients try to find another resource. If they fail
while they are in state \emph{reserved}, they never confirm the
start-up of the application to the client it has been reserved by and
the client handles this case too. If it fails when it is in
state \emph{finished}, the client has already taken into account the
fact that this resource is done executing its part of the job, so it
has no consequence on the execution of the process. The resource
goes to state \emph{unavailable} instead of \emph{available}.

The main issue is when the failure happens when the resource is
running the job. The failure is detected by a failure detector
\cite{ACT} and the machine is considered to be \emph{dead}. The client
is notified of this failure and tries to find a new machine to replace
the failed one. 

\section{Modeling the system}
\label{sec:model}

This section presents models for the two parts of the system: the
machines (section \ref{sec:model:machine}) and the reservation system
of the clients (section \ref{sec:model:reserv}). These two components
interact with each other. These interactions are detailed in section
\ref{sec:model:interaction}.

The models are presented using a finite-state formalism. Transitions
between states are represented by edges. Guards can be present on
edges, given between brackets. When the system can chose between two
transitions depending on a probability, the probability to chose a
given edge is given between parenthesis. Edges related with each
other by a probability are linked by an arc. The actions taken by a
transition is given on the corresponding edge. In particular, when a
transition can be taken only after $T$ units of time (\ie the system
must remain in a state during at least $T$ units of time), the time is
initialized before entering the state by an action on the incoming
edge ($time := 0$), and a guard on the outgoing edge sets the
condition on the time to take this transition ($time > T$).

Interactions between automata are modeled by synchronizations on
actions, in a similar way as what is presented in \cite{uppaal}. For
instance, when an automaton sends a request, the corresponding edge
contains the {\tt request!} action and the automaton receives it with
{\tt request?} on the guard of an edge.

\subsection{Model of each machine}
\label{sec:model:machine} 

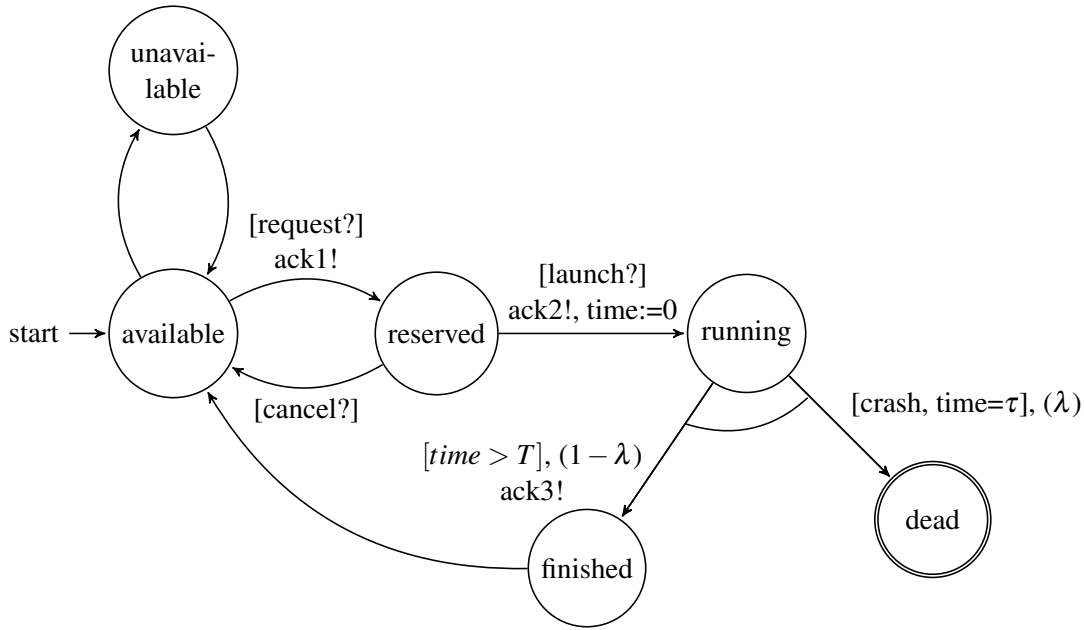
\begin{figure}[ht!b]
  \begin{center}
      \usetikzlibrary{arrows,automata}
\usetikzlibrary{shapes,snakes,backgrounds,petri,calc}
\usetikzlibrary{positioning}
\begin{tikzpicture}[->,>=stealth',shorten >=1pt,auto,node distance=3.5cm,
                    semithick, every text node part/.style={align=center}]
  \tikzstyle{every state}=[draw,text=black,minimum size=1.5cm, every text node part/.style={align=center}]

   \node[state, initial] (available) {available};
   \node[state] (unavailable) [above of=available]{unavai-\\lable};
   \node[state] (reserved) [right of=available]{reserved};
   \node[state] (running) [right = 2.5 cm of reserved]{running};
   \node[state, accepting] (dead) [below right of=running]{dead};
   \node[state] (finished) [below left = 2 cm and 1 cm of running]{finished};

\path (available) edge [bend left] node {} (unavailable)
 (unavailable) edge [bend left] node {} (available)
 (available) edge [bend left] node{[request?]\\ack1!} (reserved)
 (reserved) edge  [bend left] node{[cancel?]} (available)
 (reserved) edge  node{[launch?]\\ack2!, time:=0} (running)
 (running) edge  node{[crash, time=$\tau$], ($\lambda$)} (dead)
 (running) edge  node[left, pos=.65]{$[time>T]$, ($1-\lambda$)\\ack3!} (finished)
 (finished) edge [bend left] node {} (available)
 (running) edge  node (A) [pos=.2,xshift=-4mm] {} (finished)
 (running) edge  node [pos=.2] (B) {} (dead)
 (A) edge [-, bend right] (B)
 ;

\end{tikzpicture}
      \caption{\label{fig:machineautom} Automaton modeling the algorithm running on each machine.}
  \end{center}
\end{figure}

The states of a machine are represented in the automaton depicted on
figure \ref{fig:machineautom}.
When no job is running on it, a machine is in the \emph{available}
state. For some (local) reason, such as an action from an
administrator or a local user (when the machines are unused
workstations, for instance with desktop grids), it can become
\emph{unavailable}. When the machine is in the state \emph{available}, it can 
be reserved by a client: in this case, it enters the state
\emph{reserved}. We have seen in section \ref{sec:reservation} that a client can
cancel a reservation; in this case, the machine returns to the
state \emph{available}.

Once the client has all the machines it needs to start a job, it sends
a command to all of the machines it has reserved. These machines enter
the state \emph{running}, because it is running a process from a
job. However, the machine can crash or fail during the execution. It
happens with probability $\lambda$: then, the machine enters the state
\emph{dead}. The probability is given between parenthesis on the edges
between states.

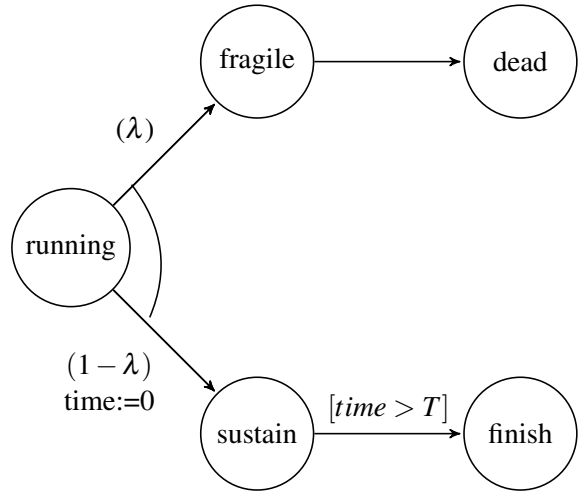
\begin{figure}[ht!b]
  \begin{center}
      \begin{tikzpicture}[->,>=stealth',shorten >=1pt,auto,node distance=3.5cm,
                    semithick, every text node part/.style={align=center}]
  \tikzstyle{every state}=[draw,text=black,minimum size=1.5cm, every text node part/.style={align=center}]

   \node[state] (running) {running};
   \node[state] (sustain) [below right of = running] {sustain};
   \node[state] (fragile) [above right of = running] {fragile};
   \node[state] (finish) [right of = sustain] {finish};
   \node[state] (dead) [right of = fragile] {dead};

\path (running) edge node[below left] {$(1-\lambda)$\\time:=0} (sustain)
 (running) edge node {($\lambda$)} (fragile)
 (fragile) edge node {} (dead)
 (sustain) edge node {$[time>T]$} (finish)
 (running) edge  node (A) [pos=.2] {} (fragile)
 (running) edge  node [pos=.2, yshift=-4mm] (B) {} (sustain)
 (A) edge [-, bend left] (B)
;
\end{tikzpicture}
      \caption{\label{fig:mort}Modeling the volatility of a resource.}
  \end{center}
\end{figure}

A more detailed model is given in figure \ref{fig:mort}. The resource
has a probability $\lambda$ of dying. If it dies, it reaches the state
\emph{fragile}. If it does not, if reaches the state
\emph{sustain}. It stays in the sustain state for (at least) $T$ time
units, representing the execution time of the process. Figure
\ref{fig:machineautom} is a more compact representation of this
subpart of the model, since the transition after the state
\emph{running} is conditioned by a probability and a time. 

Once the execution is done (\ie after a given period of
time and with probability $1-\lambda$), it enters the state
\emph{finished} and then can be \emph{available} again.

When the machine is reserved, the machine notifies the client it
acknowledges the request. Similarly, when the machine starts the
application, it acknowledges the client. At the end of the execution,
it notifies the client it is done with its local process. 

We can seen that a machine can evolve
between states, except if a failure happens and it dies. In this case,
it stays in the state \emph{dead} until an administrator performs an
action to fix the problem and restart the machine, which can take a
long time compared to the typical execution time of each transition of
the automata presented here. As a consequence, we are considering here
that the machine stays in the state \emph{dead}.

\subsection{Model of the reservation system}
\label{sec:model:reserv}

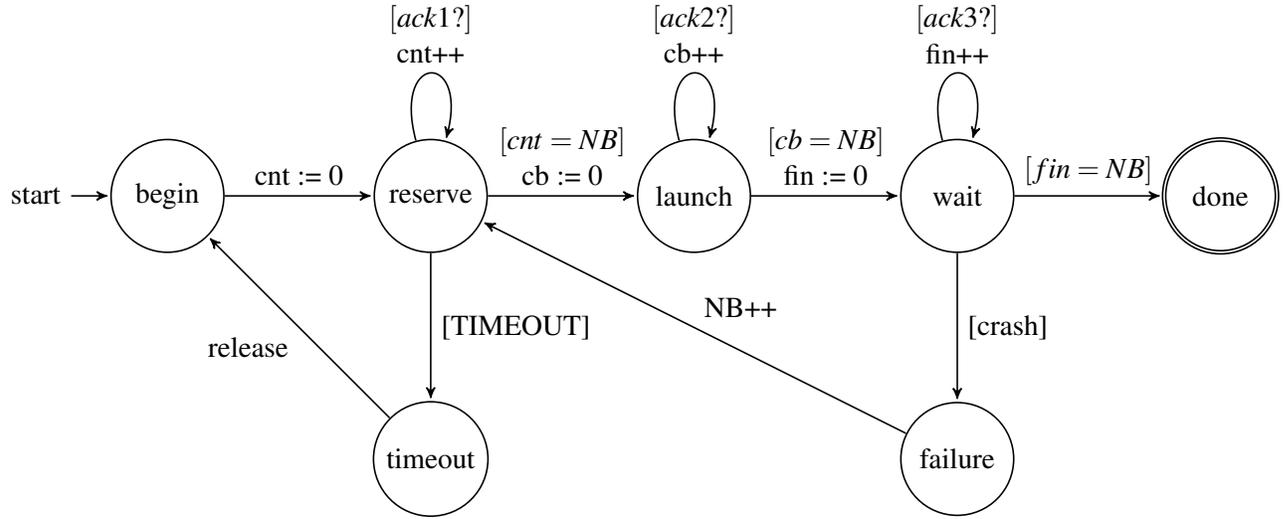
\begin{figure}[ht!b]
  \begin{center}
    \usetikzlibrary{arrows,automata}
\usetikzlibrary{shapes,snakes,backgrounds,petri,calc}
\usetikzlibrary{positioning}
\begin{tikzpicture}[->,>=stealth',shorten >=1pt,auto,node distance=3.5cm,
                    semithick, every text node part/.style={align=center}]
  \tikzstyle{every state}=[draw,text=black,minimum size=1.5cm, every text node part/.style={align=center}]

    \node[state, initial] (begin)                               {begin};
  \node[state](reserve) [right of=begin] {reserve};
  \node[state](launch) [right of=reserve] {launch};
  \node[state](wait) [right of=launch] {wait};
  \node[state, accepting](done) [right of=wait] {done};
  \node[state](timeout) [below of=reserve]{timeout};
  \node[state](failure) [below of=wait]{failure};

\path (begin) edge              node {cnt := 0} (reserve)
                (reserve)  edge [loop above] node {$[ack1?]$\\cnt++} (reserve)
				(reserve) edge node { $[cnt = NB]$\\cb := 0} (launch)
				(launch) edge [loop above] node {$[ack2?]$\\cb++} (launch)
				(launch) edge node {$[cb = NB]$\\fin := 0} (wait)
                (wait) edge [loop above] node {$[ack3?]$\\fin++} (wait)
               (wait) edge node {$[fin = NB]$} (done)
               (reserve) edge node {[TIMEOUT]}(timeout)
               (timeout) edge node {release} (begin)
               (wait) edge node {[crash]} (failure)
               (failure) edge node [above right] {NB++} (reserve);

\end{tikzpicture}
    \caption{\label{fig:automresa}Automaton modeling the reservation
      system (\emph{wait} semantics)}
  \end{center}
\end{figure}

The reservation system is modeled by the automaton given on figure
\ref{fig:automresa}. It uses counters. Initially, the system is in
state \emph{begin}. A counter is used to count the number of machines
that have acknowledged the reservation; it is initialized to zero when
the system transitions from the state \emph{begin} to the
state \emph{reserve}. Each time a machine acknowledges the
reservation, the counter is incremented. When the required number of
machines have acknowledged ($NB$ \emph{OK} answers), the automaton
reaches the state \emph{launch}. In a similar way, it counts the
number of machines that have acknowledged application start-up. Once
they all have answered, the automaton reaches the state \emph{wait},
until all the machines are done.

The automaton presented here implements the \emph{wait} semantics. If
not enough machines can be reserved, the system releases the machines
and returns to the initial state.

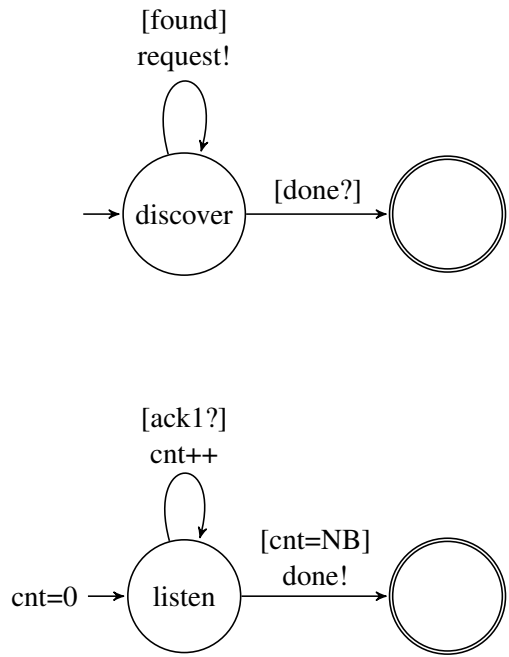
\begin{figure}[ht!b]
  \begin{center}
    \usetikzlibrary{arrows,automata}
\usetikzlibrary{shapes,snakes,backgrounds,petri,calc}
\usetikzlibrary{positioning}

\begin{tikzpicture}[->,>=stealth',shorten >=1pt,auto,node distance=3.5cm,
                    semithick, every text node part/.style={align=center}]
  \tikzstyle{every state}=[draw,text=black,minimum size=1.5cm, every text node part/.style={align=center}]

\tikzset{initial text={}}
    \node[state, initial] (discover)                               {discover};
  \node[state, accepting](done) [right of=discover] {};

\path (discover) edge              node {[done?]} (done)
				(discover) edge [loop above] node {[found]\\request!} (discover);

\tikzset{initial text={cnt=0}}

    \node[state, initial] (listen)[below=of discover]     {listen};
  \node[state, accepting](done2) [right of=listen] {};

\path (listen) edge              node {[cnt=NB]\\done!} (done2)
				(listen) edge [loop above] node {[ack1?]\\cnt++} (listen);

\end{tikzpicture}
    \caption{\label{fig:reserve}Detailed model for the state
      \emph{reserve} ({wait} semantics)}
  \end{center}
\end{figure}

A detailed model for the state \emph{reserve} is given in Figure
\ref{fig:reserve}. This model is made of two parts. The top part of
the model is the \emph{discovery} system of Zeroconf: the client
listens to the Zeroconf bus, and sends a request when it discovers a
new machine. In the same time, the client waits for acknowledgements
from the machines: this corresponds the lower part of the model. Every
time an acknowledgement is received, a counted is incremented. When
enough machines have answered, both automata reach the final state by
synchronizing on the {\tt done} action.

We have seen in section \ref{sec:volatility} that resources can
fail during the execution of a job. In this case, the reservation
system is informed by the failure detector and reaches the
state \emph{failure}. Then it requests an additional resource and
starts the application on it.

\subsection{Interactions between the automata}
\label{sec:model:interaction}

We have seen in Figure \ref{fig:machineautom} that the machines need
some actions from the reservation systems and that some actions are
made to this reservation system. In a similar way, we have seen in
figure \ref{fig:automresa} that the reservation system interacts with
the machines. This set of actions forms a mini-protocol between the
two automata.

\begin{enumerate}\itemsep0em 
\item The reservation system \emph{sends a request} to the machines it
  has found on the Zeroconf bus ({\tt request} action). When a machine
  receives a request, it answers {\it OK} or {\it KO} depending on the
  state it is currently in. 
\item The available machines \emph{acknowledge} the request ({\tt
  ack1} action). For each acknowledgement it receives, the client
  increments a counter.
\item The reservation system \emph{sends a command} to the machines
  that were assigned to it ({\tt launch} action). When each machine
  receives it, it starts the command. 
\item The available machines \emph{acknowledge} the command ({\tt
  ack2} action). The client counts the number of acknowledgement it
  receives. 
\item The available machines \emph{notify} the reservation system that
  the execution of its local process is done ({\tt ack3} action). The
  client counts the number of acknowledgement it receives and, when it
  has received all of them, the execution is done. 
\end{enumerate}

\section{Parameter synthesis and verification of the system}
\label{sec:params}

This system contains many parameters, and its behavior depends on
these parameters. Parameter synthesis can be useful to verify is
behavior under different parameters. An exploration of the behavior of
such as system, for instance using behavioral cartography
\cite{AF-rp10}, can give the ranges of parameters for the system to
behave as expected. 

\subsection{Expected behavior}
\label{sec:params:behavior}

The system must have several properties. The \emph{soundness} of the
system \cite{Vanderaalst97} (option to complete, proper
  completion and no dead transitions) was verified in \cite{CEK15, arxivmodel}: in
the absence of failures, all the jobs are executed and complete. If
resources can fail, there can be too many failures, in which case the
jobs that need more resources than there are surviving resources
cannot be executed. In this case, it was verified that \emph{there
  exists} an execution in which all the jobs complete.

More specifically, it is also necessary to ensure \emph{exclusive
  access} of the processes of all the jobs on the resources. It is one
of the expected properties of this system: when a process of a job is
executed on a resource, this resource must not be attributed to any
other job. This property is important for instance for computation
resources (computation nodes, GPUs...): if a node runs more processes
than the number of cores it has, the processes need to share the
execution time. This situation is called \emph{oversubscription}. 
This property was also verified using Petri nets in \cite{CEK15,
  arxivmodel}. 

Timed models such as timed Petri nets \cite{arxivmodel} give a more
precise idea of the behavior of the system in terms of, for example,
deadlines. Verification techniques on such models can provide
properties such as ``in $T$ time units, all the jobs have
completed''. 

Parameter synthesis is particularly important to have a more precise
idea of these properties. For instance, it can give precise completion
time for a range of execution times. It is highly important for
critical systems for example, to verify what is doable and under which
conditions. In particular, one can want to make sure that for every
possible execution, all the jobs complete before a certain number of
time units. Parameter synthesis would help dimensioning the system
(number of resources, number of jobs to put in the system) to be sure
that the system behaves as required. 

\subsection{Parameters of the model}
\label{sec:params:synthesis}

We have seen in sections \ref{sec:model:machine} and
\ref{sec:model:reserv} that the model depends on several parameters:
\begin{itemize}\itemsep0em 
\item The number of machines in the system;
\item The number of clients issuing requests in the system;
\item The number of resources asked by each client;
\item The execution time of the processes of each job;
\item The failure probability of each resource;
\item In the \emph{wait} semantics, the value of the timeout;
\item In the \emph{fail} semantics, the time after which the request
  is issued again.
\end{itemize}

The number of resources asked by each client is specific for each
client. For instance, a client may ask for three machines and another
one may ask for six machines. The execution time is assumed to be
roughly the same for all the processes of a given job but not
necessarily the same. As a consequence, machines spend some time in
the state \emph{finished}, but this time is generally small compared
to the time spent in the state \emph{running}.

However, if a resource has crashed during the execution of a process,
the failed process must be re-executed, possibly from the beginning
(some applications include a failure-recovery protocol that may reduce
the re-execution time). Therefore, at the end of the execution of this
job, the other resources used by the job are waiting for it in the
state \emph{finished}. 

Therefore, the failure rate is very important to compute a likelihood
of execution time. For instance, one can expect a result like ``There
is a likelihood of 50\% that all the applications will be done after
$N$ time units, 25\% after $2N$ time units, 15\% after $3N$ time units
and 10\% that machines will crash too often for the applications to
complete''.

Besides, keeping some parameters unknown would allow to dimension the
system with respect to some requirements. For instance, for a given
number of resources, how many jobs can be executed and finish on time?
Or the other way around, for a given number of jobs, how many 
resources does the system need to have to make sure that all the jobs
will be executed and finish on time?

\section{Conclusion}
\label{sec:conclu}

In this paper, we have presented a distributed algorithm for resource
discovery and reservation. This algorithm was verified using model
checking techniques (P/T Petri nets, Colored Petri nets and Timed
Petri nets) in a previous paper, but these tools did not take into
account the fact that the system contains several parameters,
including on times and probabilities.

Parameter synthesis techniques such as behavioral cartography would
permit to exhibit values for these parameters that would guarantee
that the system follows a certain behavior. However, the complexity of
the model, which is in the same time parametric, timed and
probabilistic, makes it challenging for current parameter synthesis
techniques.

Therefore, we hope that the system presented and modeled in this paper
will be a challenge for the parameter synthesis community to find
fitting parameter synthesis, maybe by hybridizing or combining several
existing techniques, or developing new, specific ones.

One approach would be to consider parts of the problem as smaller
instances and to decompose it as models of increasing complexity. A
finite-state model, with no probability, can be used to perform a
worst-case analysis. A real-time model, with parameterized delays,
timeouts and number of processes and resources but no probability,
can be used to do a verification and quantification of the behavior of
the system without failures. Last, the full model allows to perform
verification and behavior analysis of the complete system.

Moreover, some assumptions can be made to perform a worst-case analysis
and verify some properties. For instance, the failure probability can
be simplified by a global failure rate, such as ``at most 10\% of the
resources fail during the execution of a job''. Under this assumption,
the model loses its probabilistic nature. 

\bibliographystyle{eptcs}
\bibliography{qurd}
\end{document}